\documentclass[12pt]{article} 
\usepackage[utf8]{luainputenc}
\usepackage[english]{babel}

\usepackage{soul}
\usepackage{amsmath}
\usepackage{amsthm}
\usepackage{amscd}
\usepackage{amssymb}
\usepackage{dsfont}
\usepackage{bm}
\usepackage{caption}
\usepackage{leftidx}
\usepackage[makeroom]{cancel}
\usepackage{indentfirst}
\usepackage{hyperref}
\usepackage{tikz}
\usepackage[com pat=1.1.0]{tikz-feynman}
\usepackage{subfig}
\usepackage{stackengine}
\usepackage{enumitem}
\usepackage{courier}
\usepackage{graphicx}
\usepackage{mathtools}
\usepackage{tikz-cd}
\usepackage{euscript}
\usepackage[titletoc,title]{appendix}

\usepackage{adjustbox}
\usepackage{mathtools}
\usepackage{tensor}
\usepackage{multirow}
\usetikzlibrary{matrix,arrows,decorations.pathmorphing}
\usepackage{comment}
\usepackage{geometry}
\newcommand {\unit} {\hat{\mathds{1}}}

\DeclareMathOperator{\Tr}{Tr\,  }

\DeclareMathOperator{\const}{const\,  }

\newcommand{\bma} {\begin{pmatrix}}
\newcommand{\ema} {\end{pmatrix}}

\newcommand{\iu}{{i\mkern1mu}}

\renewcommand{\d}[1]{\ensuremath{\operatorname{d}\!{#1}}}
    
\newtheoremstyle{dotless}{}{}{\itshape}{}{\bfseries}{}{ }{}
\theoremstyle{dotless}

\newcommand{\s}{\nobreak\hspace{.11em}\nobreak}
\newcommand{\beq}{\begin{equation}}
\newcommand{\eeq}{\end{equation}}
\newcommand{\beqn}{\begin{equation}\left.\begin{aligned}}
\newcommand{\eeqn}{\end{aligned}\right.\end{equation}}

\newcommand{\sQ}{\EuScript{Q}}

\newcommand{\bpma} {\begin{pmatrix*}[r]}
\newcommand{\epma} {\end{pmatrix*}}
\newcommand{\pmat} [4] {\begin{pmatrix*}[r] #1 & #2 \\ #3 & #4 \end{pmatrix*}}

\newcommand{\pcolt} [2] {\begin{pmatrix*}[r] #1 \\ #2\end{pmatrix*}}

\newcommand{\ie}{{\mbox{i.e.} }}

\newcommand{\BB}{\mathcal{B}{}}

\newcommand{\beqS}{\begin{equation}}
\newcommand{\eeqS}{\end{equation}}
\newcommand{\beqSn}{\begin{eqnarray}}
\newcommand{\eeqSn}{\end{eqnarray}}

\renewcommand{\L}{\mathcal{L}}
\newcommand{\D}{\mathcal{D}}

\newcommand{\Dq}{\D q}
\newcommand{\Phib}{{\Phi^\dagger}}
\newcommand{\phib}{{\phi^\dagger}}

\newcommand{\qb}{\overline{q}{}}

\newcommand{\CP}{\href{http://lurkmore.to/\%D0\%A1\%D0\%A0}{\mathbb{CP}}}
\renewcommand{\L}{\mathcal{L}}

\newcommand{\alphab}{\bar{\alpha}}
\newcommand{\betab}{\bar{\beta}}
\newcommand{\gammab}{\bar{\gamma}}
\newcommand{\deltab}{\bar{\delta}}

\newcommand{\Ab}{{A^\dagger}{}}
\newcommand{\Bb}{{B^\dagger}{}}
\newcommand{\BBb}{{\mathcal{B}^\dagger}{}}

\newcommand{\ib}{\bar{i}}
\newcommand{\jb}{\bar{j}}

\newcommand{\lb}{\bar{l}}

\newcommand{\nb}{\bar{n}}

\newcommand{\psib}{\overline{\psi}}

\usepackage[autostyle]{csquotes}
\usepackage[
    backend=biber, 
    sorting=none, 
    style=numeric-comp,
    hyperref=true,
    doi=false,
    isbn=false,
    giveninits=true,
    maxbibnames=99
]{biblatex}
\renewbibmacro{in:}{}
\AtEveryBibitem{%
  \clearfield{note}%
}

\DeclareFieldFormat[article]{volume}{\mkbibbold{#1}} 
\DeclareFieldFormat[article]{number}{(#1)} 
\DeclareFieldFormat{url}{\newline #1} 
\DeclareFieldFormat{doi}{\newline #1} 

\usepackage{authblk}
\makeatletter
\renewcommand\AB@authnote[1]{\rlap{\textsuperscript{\normalfont#1}}}

\makeatother
\author[1,2]{Michael Kreshchuk\s}
\author[2]{Evgeniy Kurianovych\s}
\author[2,3]{Mikhail Shifman\s}
\affil[1]{{\small
~Department of Physics, Tufts University, Medford, MA, 02155, USA
}}
\affil[2]{{\small
~School of Physics and Astronomy, University of Minnesota, Minneapolis, MN, 55455, USA}
}
\affil[3]{{\small
~William I. Fine Theoretical Physics Institute, University of Minnesota, Minneapolis, MN, 55455, USA}
}
\title{
{\Large{\textbf{
On Grassmannian Heterotic Sigma Model
}}}}
\date{}
\makeatother
\begin{document}
\begin{titlepage}
\begin{flushright}
FTPI-MINN-18/24, UMN-TH-3809/8
\end{flushright}
{\let\newpage\relax\maketitle}
\maketitle
\begin{abstract}
We study the non-minimal supersymmetric heterotically deformed $\mathcal{N}=(0,2)$ sigma model with the Grassmannian  target space $\mathcal{G}_{M,N}$. To develop the appropriate superfield formalism, we begin with a simplified model with flat target space, find its beta function up to two loops, and prove a non-renormalization theorem. Then we generalize the results to the full model with the Grassmannian target space. Using the geometric formulation, we calculate the beta functions and discuss the 't Hooft and Veneziano limits.
\end{abstract}
\end{titlepage}

\tableofcontents

\section*{Introduction}

Sigma-models have a long history  dating to 1960. \cite{GellMann:1960np}.
In $1975$, Polyakov~\cite{Polyakov:1975rr} was the first to observe that the $\mathbb{O}(3)$ sigma model is asymptotically free and provides a laboratory for modelling 4$D$ gauge theories, allowing one to study similar phenomena in a simplified setting. Heterotically deformed supersymmetric models are of particular interest due to their close connection to the world sheet theories on vortex strings supported in the ${\mathcal{N}=1}$ Super-Yang-Mills in four dimensions \cite{Shifman:2008wv}. Additionally, such models are also interesting from the mathematical viewpoint~\cite{Witten:1993xi,Bourdeau:1994je,Stolz,Bak:2006qk}. In the recent years heterotic sigma-models attracted a significant attention \cite{Adams:2003zy,Witten:2005px,Shifman:2008kj,Tan:2008mi,Yagi:2010tp,Koroteev:2010gt,Koroteev:2010ct,Cui:2010si,Cui:2011rz,Cui:2011uw,Melnikov:2012hk,Gadde:2013lxa,Gadde:2014ppa,Jia:2014ffa,Chen:2014efa,Adam,Chen:2015dti,Chen:2015xda,Dunne:2015ywa} (for a review see \cite{Shifman:2014jba}).

The most widely considered target spaces of the sigma models studied in the literature are ${\mathbb{S}^{N}=\mathbb{SO}(N+1)/\mathbb{SO}(N)}$~\cite{Polyakov:1975rr} and ${\mathbb{CP}(N)=\mathbb{SU}(N+1)/\mathbb{SU}(N)\times \mathbb{U}(1)}$~\cite{Eichenherr:1978qa,Golo:1978dd,Cremmer:1978bh}. Their supersymmetric versions were introduced in~\cite{Cremmer:1978bh} and~\cite{Witten:1977xn,DiVecchia:1977nxl,Novikov:1984ac}, respectively. These arise naturally as the world-sheet effective low-energy theories of the non-Abelian string solutions. In this paper, we turn to the generalization of $\mathbb{CP}(N)$, the Grassmannian target space~\cite{Pisarski:1979gw,vanHolten:1983di,Morozov:1984ad,Abdalla:1984en},
\begin{equation}
{\cal G}_{M,N}=\frac{\mathbb{SU}(N+M)}{\mathbb{SU}(N)\times\mathbb{SU}(M)\times
\mathbb{U}(1)}\quad.
\label{GrMan}
\end{equation}

Since the Grassmannian is a K\"ahler manifold, the Lagrangian of the $(2,2)$ model can be written as \begin{equation}
    \mathcal{L}_{(2,2)} = \int \d{}^4 \theta K(\Phi,\Phib) \quad,
    \label{InitialKahler}
\end{equation}
where $K(\Phi,\Phib)$ is the K\"ahler potential. In our paper, we consider the so-called non-minimal model in which supersymmetry is broken down to $(0,2)$ by the addition of an extra deformation term to the Lagrangian \eqref{InitialKahler}. In this way, one ends up with two types of interactions and two corresponding coupling constants. The interaction of the first type is owing to geometry of the target space, while the interaction of the second type is due to the heterotic deformation.

In Section \ref{section:flat}, we develop the superfield formalism and study the limit of the heterotic deformation assumed to be much stronger than the interaction due to the target-space geometry. To this end, we consider the simplified flat target space model, a straightforward generalization of the case considered in~\cite{Cui:2011rz}. In Section \ref{section:nonrenorm}, we prove a non-renormalization theorem for this model. In Section \ref{section:full}, we consider the full model, with both the curved target space and the heterotic deformation. There are two ways of analyzing interactions induced by the target space constraint: 
(a) by approaching the curved geometry directly, and (b) by introducing extra gauge fields. In the current work we shall rely on the former approach. For its implementation, several basic tools from geometry of the Grassmannian will be required. A brief review of this machinery is also provided in Section \ref{section:full}. In Section \ref{section:beta}, we write down the beta functions of the full model, while their 't Hooft and Veneziano limits are discussed in Section \ref{section:largeN}. Our notation, as well as the background field method, are reviewed in the appendices.

\section{Flat target space model}\label{section:flat}

The non-minimal heterotic $(0,2)$ model combines the original $(2,2)$ model and its deformation that partly breaks supersymmetry. Before exploring their interplay, let us take a separate look at the two parts~---~the interaction imposed by geometry and the interaction generated by deformation.
The case of the undeformed $(2,2)$ model is well studied in the literature~\cite{Morozov:1984ad,Perelomov:1987va,Abdalla:1984en,Perelomov:1989im}. In this section we will start from exploring the opposite limit~--- the heterotically deformed flat target space model. Since its target space is just the complex linear $\mathbb{C}^{MN}$, we'll call this model linearized. It is a generalization of the case considered in~\cite{Cui:2011rz}, and will be used to develop the appropriate $(0,2)$ formalism. This theory corresponds to the limit of the vanishing target-space curvature of the full model, or, in other words, to the interaction caused by the heterotic deformation  much stronger than the interaction arising from geometry.
For this linearized model with a trivial metric, we shall study the heterotic deformation and shall discuss the running of the coupling constant. Then, we shall generalize the results to the full model~\eqref{Lfull}.

The complex dimension of the Grassmannian manifold is $MN$. Accordingly, we introduce $MN$ chiral superfields labeled by two indices:
\beq\label{mama}
\Phi^{i \alpha}
=\phi^{i \alpha}+
\sqrt{2}\theta\psi^{i \alpha}+\theta^2 F^{i \alpha}
\quad,\qquad  n=1\ldots N,\quad \alpha=1\ldots M \,.
\eeq
In terms of the $(0,2)$ superfields, $\Phi^{i,\alpha}$ can be written as
\begin{equation}
\begin{multlined}
\Phi^{i \alpha}(x_{R}+2i\theta_{R}^{\dagger}\theta_{R},x_{L}-2i\theta_{L}^{\dagger}\theta_{L},\theta_{R},\theta_{L})
\\=A^{i \alpha}(x_{R}+2i\theta_{R}^\dagger\theta_{R},x_{L}-2i\theta_{L}^{\dagger}\theta_{L},\theta_{R})
+\sqrt{2}\,\theta_{L} B^{i \alpha}(x_{R}+2i\theta_{R}^\dagger\theta_{R},x_{L},\theta_{R})\quad.
\end{multlined}
\label{decomp}    
\end{equation}
The superfield $A^{i \alpha}$ represents the chiral supermultiplet; on mass shell it consists of the scalar field and the left-moving fermion,
\begin{alignat}{9}
A^{i \alpha}&=\phi^{i \alpha}(x_{R}+2i\theta^\dagger_R\theta_R,\,x_{L})&&+\sqrt{2}\,\theta_R\,\psi^{i \alpha}_L(x_{R}+2i\theta^\dagger_R\theta_R,\, x_{L}) \quad&&,   
\label{Afield}
\intertext{where}
&\qquad x_L = \dfrac{1}{2}(x^0 + x^1)
\quad&&,\qquad
x_R = \dfrac{1}{2}(x^0 - x^1)\quad.
\intertext{
The field $B^{i \alpha}$ describes the Fermi supermultiplet which on mass shell contains only the right-moving fermion
($F^{i \alpha}$ is an auxiliary field),
}
B^{i \alpha}&=\psi^{i \alpha}_R(x_{R}+2i\theta^\dagger_R\theta_R,\, x_{L})&&+\sqrt{2}\,\theta_R F^{i \alpha}_\psi(x_{R}+2i\theta^\dagger_R\theta_R,\, x_{L}) \quad&&.
\label{Bfield}
\intertext{
To break supersymmetry down to $(0,2)$, we introduce another supermultiplet~$\BB$:
}
{\mathcal B}&=\zeta_R(x_{R}+2i\theta^\dagger_R\theta_R,\,x_{L})&&+\sqrt{2}\,\theta_R\,F_\zeta(x_{R}+2i\theta^\dagger_R\theta_R,\, x_{L})\quad.
\label{Bzeta}
\end{alignat}
It has no target space indices.

\begin{figure}\captionsetup{font=footnotesize}
\centering
  \subfloat[]{%
    \includegraphics[width=0.3\textwidth]{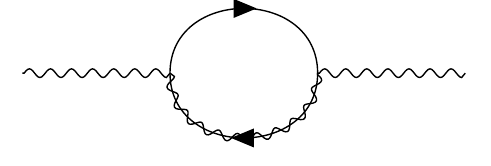}}
  \subfloat[]{%
    \includegraphics[width=0.3\textwidth]{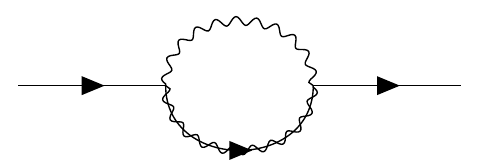}}
  \subfloat[]{%
    \includegraphics[width=0.3\textwidth]{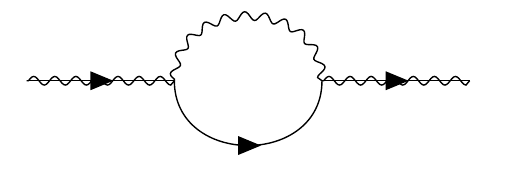}} 
  \caption{one loop corrections. Wavy line is $A$, solid line is $B$ and composition of the wavy and solid lines is $\BB$.}
  \label{OneLoopDiagrams}
\end{figure}

In the ${(0,2)}$ formalism, the Lagrangian of the linearized model acquires the form of~\footnote{~In this section, due to the flatness of the target space, the position of fields' indices is not of importance. For the same reason, here we do not use the barred indices either.}
\begin{equation}
\label{Llin}
\begin{alignedat}{9}
    \mathcal{L}_{\text{flat}} = \dfrac{1}{2} \int \d{}^2 \theta_R
    \biggl[
    \dfrac{1}{2}Z_A &\left( \iu \Ab_{i\alpha} \partial_R A^{i\alpha} \right)
    + Z_{B} \Bb_{i \alpha} B^{i\alpha}
    \\&+ Z_{\BB} \BBb \BB
    -Z_\gamma \left( \gamma \BB B^{i\alpha} \Ab_{i\alpha} + \text{H.c.}  \right)
    \biggr] \quad.
\end{alignedat}\end{equation}

Here we have already introduced the $Z$-factors. The last term is the one breaking the supersymmetry down to $(0,2)$. We see that such a model corresponds to $MN$ copies of the linearized $\CP(1)$ considered in~\cite{Cui:2011rz}, with $A$ and $B$ fields interacting only via the deformation field.

After eliminating the auxilary fields $F^{i \alpha}_\psi$ and $F_\zeta$, equation~\eqref{Llin} reads in components as:
\begin{equation}\begin{multlined}
    \mathcal{L}_{\text{flat}} = 
    \partial^{\mu} \phi^{\dagger}_{i\alpha} \partial_\mu \phi^{i\alpha}
    + \iu \overline{\psi}_{i\alpha} \cancel{\partial} \psi^{i\alpha}
    + \iu \zeta^{\dagger}_R \partial_L \zeta_R
    + \left(\gamma \zeta_R \psi_R^{i\alpha} \partial_L \phi^{\dagger}_{i\alpha}
    +\text{H.c.}\right)
    \\+\gamma^2 \left(\zeta^\dagger_R \zeta_R\right)\left(\psi^\dagger_{L\s i\alpha} \psi_L^{i\alpha}\right)
    +\gamma^2 \left(\psi^\dagger_{R\s i\alpha} \psi_R^{i\alpha}\right)\left(\psi^\dagger_{L\s i\alpha} \psi_L^{i\alpha}\right)
    \quad.
\end{multlined}\end{equation}

\begin{figure}\captionsetup{font=footnotesize}
\centering
  \subfloat[]{%
    \includegraphics[width=0.35\textwidth]{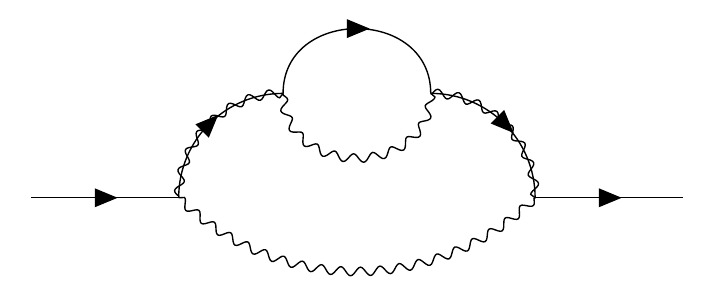}}
  \subfloat[]{%
    \includegraphics[width=0.35\textwidth]{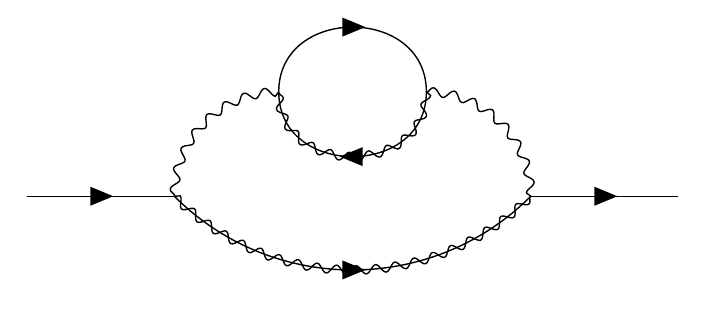}}
  \subfloat[]{%
    \includegraphics[width=0.25\textwidth]{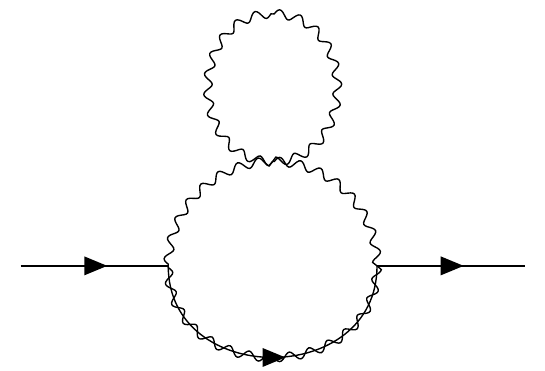}} \\
  \subfloat[]{%
    \includegraphics[width=0.35\textwidth]{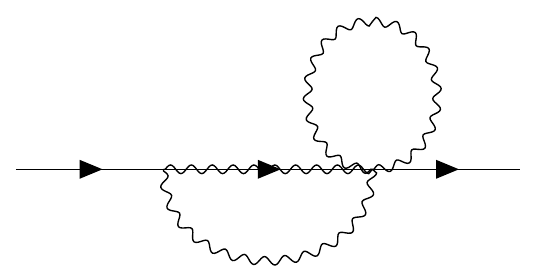}}
  \subfloat[]{%
    \includegraphics[width=0.35\textwidth]{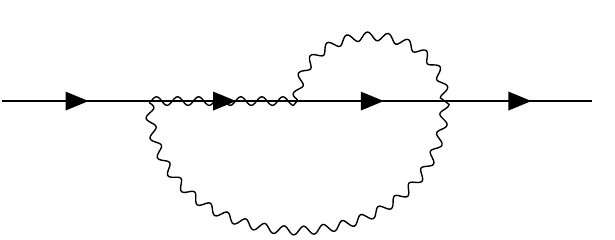}}
  \subfloat[]{%
    \includegraphics[width=0.2\textwidth]{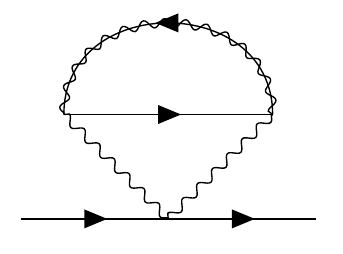}} 
  \caption{Two loop corrections for $B$.}
  \label{Btulup}
\end{figure}

\begin{figure}\captionsetup{font=footnotesize}
\centering
  \subfloat[]{%
    \includegraphics[width=0.3\textwidth]{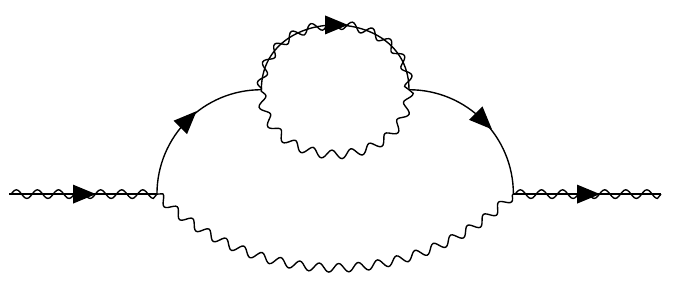}}
  \subfloat[]{%
    \includegraphics[width=0.3\textwidth]{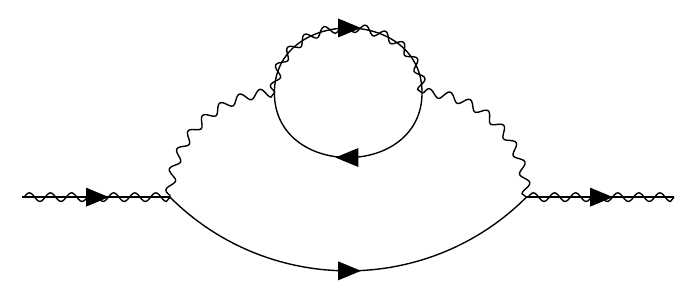}}
  \subfloat[]{%
    \includegraphics[width=0.3\textwidth]{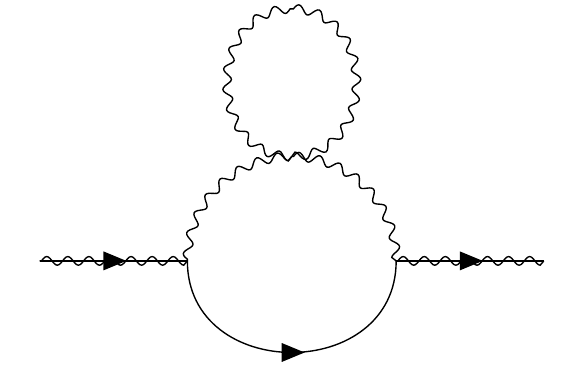}} \\
  \subfloat[]{%
    \includegraphics[width=0.25\textwidth]{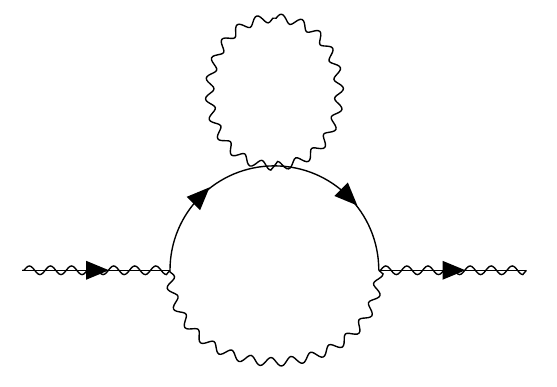}}
  \subfloat[]{%
    \includegraphics[width=0.3\textwidth]{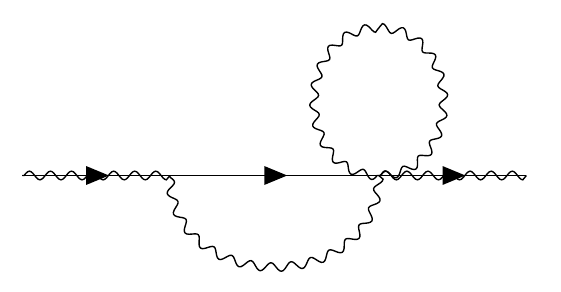}}
  \subfloat[]{%
    \includegraphics[width=0.3\textwidth]{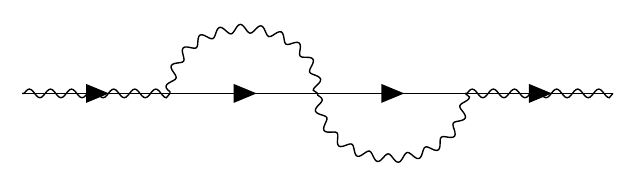}} 
  \caption{Two loop corrections for $\BB$.}
  \label{BBtulup}
\end{figure}

The diagrams for one-loop wave function renormalization are shown in Figure~\ref{OneLoopDiagrams}, for two-loop~--- in \mbox{Fig.~\ref{Btulup} (a, b)} and \mbox{Fig. \ref{BBtulup} (a, b)}. They give:
\begin{subequations}\label{zfa}\begin{alignat}{9}
    Z_{B} &= 1 + \iu \gamma^2 I + \dfrac{1}{2}  M\s N\gamma^ 4I^2 \quad&&,\\
    Z_\BB &= 1 + \iu M\s N \gamma^2 I + \dfrac{1}{2} M\s N \gamma^4 I^2 \quad&&,
\end{alignat}\end{subequations}
where
\begin{equation}\begin{multlined}
    I = \int \dfrac{\d{}^{ 2-\epsilon}\s p}{(2\pi)^{2-\epsilon}} \dfrac{1}{p^2-\mu^2}
    = -\dfrac{\iu}{2\pi\epsilon} + O(\epsilon^0)= -
    \dfrac{\iu}{2\pi} \log \left( \dfrac{M_{\text{uv}}}{\mu}  \right) \quad.
\end{multlined}\end{equation}
At the one-loop level, there are no diagrams contributing to renormalization of $\gamma$, and so the beta function is determined solely by the $Z$ factors~\eqref{zfa}.
Consequently, two two-loop beta function for $\gamma^2$ is
\begin{equation}\label{betaf}
    \beta(\gamma^2) \frac{\partial \gamma^2}{\partial \log \mu^2}= \dfrac{(M\s N+1)\gamma^4}{2\pi}
    \quad.
\end{equation}

As was shown in~\cite{Cui:2011rz}, not only the beta function for the coupling $\gamma$ is exact to all orders in perturbation theory -- it does not receive 
non-perturbative corrections either.
The superfield $A$ is also not renormalized due to the theorem proven in~\cite{Cui:2011rz}.

The positivity of the beta function~\eqref{betaf} implies the existence of the Landau pole, which indicates that the model should be considered as an effective low-energy theory having some UV completion. We shall shortly see that under certain conditions the full model exhibits the same type of behavior.

\section{Non-renormalization}\label{section:nonrenorm}

We now prove a version of the non-renormalization theorem for the interaction term and the kinetic term of the $A$ field. To this end, we use the symmetries and analytic properties analogous to those from~\cite{Cui:2011rz}, and follow the way of reasoning similar to that in \cite{Seiberg:1993vc}. Our theorem is to be valid to all orders in the perturbation theory and, most importantly, non-perturbatively as well.

Let us take a look at the $R$-symmetry. Most of the terms in the Lagrangian are neutral combinations of the type $A^\dagger A$, $B^\dagger B$ and $\BB^\dagger \BB$. The only term we have to care about is 
\beq
\gamma\BB BA^\dagger\quad.
\eeq
For this term, we are free to choose three independent $\mathbb{U}(1)$ phases, while the fourth one will be defined by the condition of the overall neutrality. So, we have a three-dimensional space whose sample basis is provided in Table \ref{U1charges}.

\begin{table}\captionsetup{font=footnotesize}
\begin{center}
\begin{tabular}{|c|c|c |c| }
\hline
&$\mathbb{U}(1)_1$ &$\mathbb{U}(1)_2$ &$\mathbb{U}(1)_3$ \\
\hline
 $A$ &1 &1 &1\\
 \hline
 $B$ &1 &0 &0\\
 \hline
$\BB$ &0 &1 &0\\
\hline
$\gamma$ &0 &0 &1\\
\hline
\end{tabular}
\caption{$\mathbb{U}(1)$ charges of the model.}
\label{U1charges}
\end{center}
\end{table}

Assume that we have a function of $\gamma\BB B A^\dagger$ and $A^\dagger A$. The most general function of the neutral combinations can be expressed as
\beq
f\left(\frac{\gamma \BB B}{A},\gamma\BB B A^\dagger\right)\quad.
\eeq
The integral over $d^2\theta_R$ must be invariant under the linear shifts \beq
A^\dagger\rightarrow A^\dagger+a^\dagger\quad,
\label{shift}
\eeq
which requires the integrand to be a combination of holomorphic and antiholomorphic functions. Its Laurent expansion in powers of $\gamma\BB B A^\dagger$ is exhausted by the constant and linear terms:
\beq
f=f_0\left(\frac{\gamma \BB B}{A}\right)+
f_1\left(\frac{\gamma \BB B}{A}\right)\gamma\BB B A^\dagger\quad.
\eeq
At $\gamma=0$ the theory is free, so there should be no negative powers of $\gamma$ in the Laurent series for $f_{0, 1}$. $A$ in the denominator and the shift symmetry \eqref{shift} restrict the positive powers, so $f_{0, 1}$ should be constants. We see that there is no dependence on $A^\dagger A$; in the same way we can prove the absence of $|B|^2$, $|\BB|^2$ and $|\gamma|^2$.
After the integration over $d^2\theta_R$, $f_0$ vanishes:
\beq
\int d^2\theta_R f_0 = 0.
\eeq
Since $f_1$, being a constant, doesn't depend on $\gamma$, it has to be the coefficient from the classical Lagrangian \eqref{Llin}.

The renormalized kinetic term for $A$ will look like
\beq
FA^\dagger\partial_R A+F^\dagger A\partial_R A^\dagger\quad.
\eeq
Invariance under the shift symmetry
\beq
A\rightarrow A + a_1(t-z), \quad
A^\dagger\rightarrow A^\dagger + a_2(t-z)
\eeq
requires $Z_A$ to be 1.

\section{The full model and its geometry}\label{section:full}

The complex Grassmannian manifold is a manifold consisting of all $M$-dimensional subspaces of an $(M+N)$-dimensional complex vector space. Since a subspace is uniquely determined by its complement, it can also be treated as a manifold of $N$-dimensional subspaces, which makes ${\cal G}_{M,N}$ and ${\cal G}_{M,N}$ equivalent. The Grassmannian is a homogenious space, i.e. a space that can be represented as a quotient of a group, acting transitively on a manifold, over the stabilizer of a certain element. In the present case, $\mathbb{SU}(N+M)$ acts on $\mathbb{C}^{N+M}$ whose $M$-dimensional subspace is invariant under its rotations, given by $\mathbb{SU}(M)$, and rotations of the complement, given by $\mathbb{SU}(N)$, which justifies the definition \eqref{GrMan}.
The complex Grassmannian admits a \textit{K\"ahler-Einstein metric}~--- the one which is both K\"ahlerian (\ie is defined by a single real closed ${(1,1)}$ form, the \textit{K\"ahler potential}, see equation~\eqref{Kpot} below) and is proportional to the Ricci tensor,
\beq
{R}_{i\bar{j}}=b\frac{g^2}{2}G_{i\bar{j}}\quad,
\label{GenHomRicMet}
\eeq
where $g$ is the coupling constant and $b$ is the dual Coxeter number.

The complex dimension of the Grassmannian manifold is $MN$. In terms of the $MN$ chiral superfields, labeled by two indices, that were introduced above,
\beq\label{mama}
\Phi^{i \alpha}
=\phi^{i \alpha}+
\sqrt{2}\theta\psi^{i \alpha}+\theta^2 F^{i \alpha}
\quad,\qquad  n=1\ldots N,\quad \alpha=1\ldots M \quad,
\eeq
a generic undeformed ${\cal N}=(2,2)$ model the Lagrangian can be written as
\begin{equation}\label{KK}
    \mathcal{L}_{(2,2)} = \int \d{}^4 \theta K(\Phi,\Phib) \quad,
\end{equation}
where $K(\Phi,\Phib)$ is the K\"ahler potential, which depends on the chiral and antichiral fields.

Since interactions in the undeformed model are caused by geometry, we will now review some details about geometric structure of the target space. The K\"ahler potential of the Grassmannian manifold is given by
\begin{equation}\label{Kpot}
    K = \dfrac{2}{g^2} \Tr \ln (\delta^{n\bar{m}} + \Phi^{n\gamma}\Phib^{\bar{\gamma} \bar{m}}) \quad,
\end{equation}
where the trace is taken over the Latin indices. Obviously, we could also define ${K = \Tr \ln (\delta^{\alpha\bar{\beta}} + \Phi^{n\alpha}\Phib^{\bar{\beta}\bar{n}})}$ and take the trace with respect to the Greek indices. The subscripts are reserved for lowering with the aid of the metric tensor.
The first derivatives of the K\"ahler potential have the form of:
\begin{equation}\begin{alignedat}{9}
    K^{\betab \jb} &\equiv
    \dfrac{\partial}{\partial \Phib^{\betab \jb}} K
    =  \dfrac{2}{g^2}  \Phi^{n\beta}
        [(\unit_N + \Phi \s \Phib)^{-1}]^{\jb n}
    \quad&&,\\
        \overline{K}^{\alpha i} &\equiv
    \dfrac{\partial}{\partial \Phi^{\alpha i}} K
    = \dfrac{2}{g^2}  \Phib^{\nb\alphab}
        [(\unit_N + \Phi \s \Phib)^{-1}]^{\ib n}
        \quad&&.
\end{alignedat}\end{equation}

The K\"ahler metric is obtained as follows:
\begin{gather}\begin{multlined}
    G_{i\bar{j} \alpha \bar{\beta}}
    =
    \dfrac{\partial}{\partial \Phi^{i\alpha}}
    \dfrac{\partial}{\partial \Phib^{\bar{\beta}\bar{j}}}
    K = \dfrac{2}{g^2}\Tr \biggl\{
    \delta^{ni}\delta^{\alpha\bar{\beta}}[(\unit_N + \Phi \s \Phib)^{-1}]^{\bar{j} m}
     \\- \Phi^{n\beta}
    [(\unit_N + \Phi \s \Phib)^{-1}]^{\bar{j} i}
    \Phib^{\bar{\alpha} \bar{l}}
    [(\unit_N + \Phi \s \Phib)^{-1}]^{\bar{l} m}
    \biggr\}
    \\= \dfrac{2}{g^2}[(\unit_N + \Phi \s \Phib)^{-1}]^{\ib j} [(\unit_M + \Phi \s \Phib)^{-1}]^{\alphab \beta} \quad.
\end{multlined}\end{gather}

The small-$\Phi$ expansion of the metric, which is to be used in the background-field method, has the form:
\begin{alignat}{9}
    G_{i\jb\alpha\betab} =
    \dfrac{2}{g^2}\left[
    \delta^{i\jb}\delta^{\alpha\betab}
    - \delta^{\alpha\betab}\Phi^{j\gamma} \Phib^{\gammab\ib}
    - \delta^{i \jb}\Phi^{n\beta} \Phib^{\alphab n}
    \right] + \ldots
\end{alignat}
where the dots stand for the higher-order terms.

The Ricci tensor is proportional to the metric,
\begin{equation}
 {R}_{i\jb\alpha\betab} = \dfrac{g^2}{2}(M+N) G_{i\jb\alpha\betab} \quad,
\end{equation}
which is a particular case of \eqref{GenHomRicMet}. Further details can be found in a review paper \cite{Perelomov:1987va}, Sect. 4.2.


We are now in a position to apply these results to the full model which combines the geometric structure with the partial supersymmetry breaking. To break supersymmetry, we add another term to the Lagrangian, which is similar to the one in the previous section.

Since $\BB$ is a singlet with respect to the isometry group,
its introduction does not affect the geometry of the model. Thus, we get the following expression for the Lagrangian of the full model in the $(0,2)$ formalism: 
\begin{equation}\label{Lfull}\begin{multlined}
\mathcal{L}_{\mathcal{G}_{M,N}} =
\dfrac{1}{4} \d{}^2 \theta  \left[
\overline{K}_{\alpha i} (A, \Ab) \s
\left(\iu \partial_R A^{\alpha i} - 2 \kappa \BB B^{\alpha i}\right)
+ \text{H.c.}
\right]
\\+\dfrac{1}{2}\int \d{}^2 \theta\left[
Z_B\s G_{i\bar{j} \alpha \bar{\beta}}(A, A^\dagger)\s \Bb^{\betab \jb} B^{\alpha i}  +  Z_\BB \s\mathcal{B}^\dagger \mathcal{B}\right] \quad.
\end{multlined}\end{equation}

Next, we proceed to finding the beta functions of this model.

\medskip

\begin{figure}\captionsetup{font=footnotesize}
\centering
  \subfloat[]{%
    \includegraphics[width=0.35\textwidth]{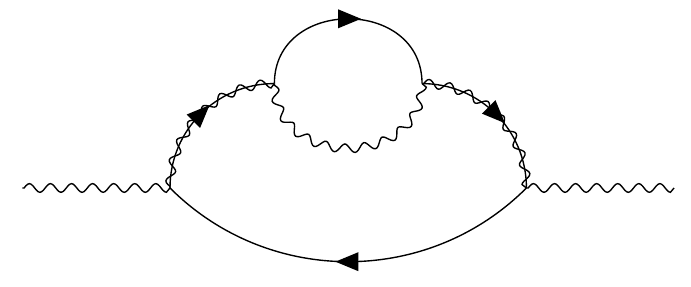}}
  \subfloat[]{%
    \includegraphics[width=0.35\textwidth]{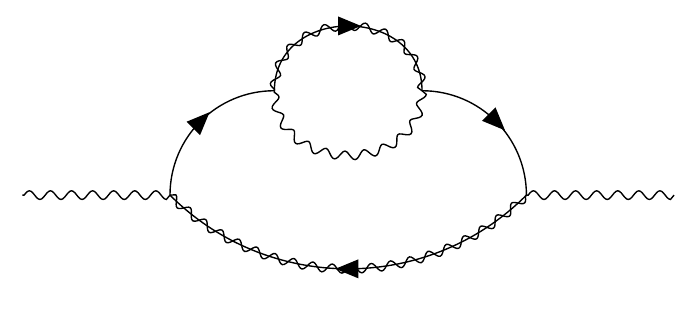}}
  \subfloat[]{%
    \includegraphics[width=0.25\textwidth]{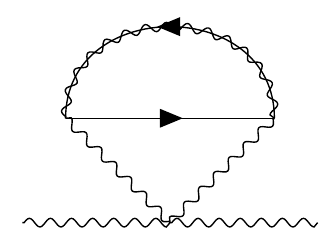}} \\
  \subfloat[]{%
    \includegraphics[width=0.3\textwidth]{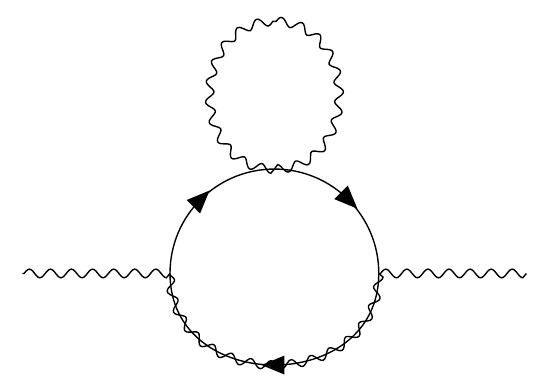}}
  \subfloat[]{%
    \includegraphics[width=0.35\textwidth]{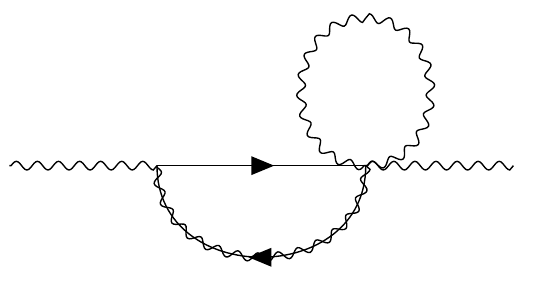}}
  \subfloat[]{%
    \includegraphics[width=0.35\textwidth]{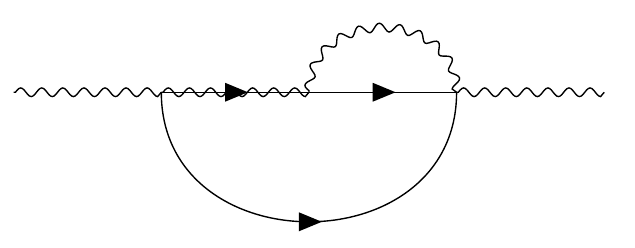}} 
  \caption{Two loop corrections for $A$. The diagrams (a) and (b) are common for both the linearized and the full models, while the remaining ones involve vertices with more than three lines which appear only in the geometric model.}
  \label{Atulup}
\end{figure}

\section{Beta functions of the full model}\label{section:beta}

In components, the Lagrangian of the full model $\eqref{Lfull}$ reads as
\begin{equation}\label{components}\begin{multlined}
\mathcal{L}_{\mathcal{G}_{M,N}} =
G_{i\bar{j} \alpha \bar{\beta}}
\left[
\partial_{L}{\phi}^{\dagger\s\bar j\bar\beta} \partial_{R}\phi^{\alpha i}
+\psi_{L}^{\dagger\s\bar j\bar\beta}\iu\nabla_{R}\s\psi_{L}^{\alpha i}
+ Z_B\s\psi_{R}^{\dagger\s\bar j\bar\beta}\s \iu\nabla_{L}\psi_{R}^{\alpha i}\right]
 \\+ Z_B\s R_{i\jb k \lb \alpha\betab\gamma\deltab}\s\psi_{L}^{\dagger\s\bar j\bar\beta}\psi_{L}^{\alpha i}\s\psi_{R}^{\dagger\s\bar l\bar\delta}\psi_{R}^{\gamma k}
+{Z_\BB}\s\zeta_R^\dagger \s i\partial_L \s \zeta_R  +
\left[\kappa\s \zeta_R  \s G_{i\bar{j} \alpha \bar{\beta}}\big( i\s\partial_{L}\phi^{\dagger\s\bar j\bar\beta} \psi_R^{\alpha i})
+{\rm H.c.}\right]
\\+\frac{|\kappa |^2}{Z_B} \zeta_R^\dagger\s \zeta_R
\big(G_{i\bar{j} \alpha \bar{\beta}}\s  \psi_L^{\dagger\s\bar j\bar\beta}  \psi_L^{\alpha i}\big)
-\s\frac{|\kappa |^2}{{Z_\BB}}
 \big(G_{i\bar{j} \alpha \bar{\beta}}\s \psi_{L}^{\dagger\s\bar j\bar\beta}\psi_{R}^{\alpha i}\big) \big(G_{k \lb \gamma \deltab}\s\psi_{R}^{\dagger\s\bar l\bar\delta}\psi_{L}^{\gamma k}\big) \quad.
\end{multlined}\end{equation}

Here $\kappa$ is the heterotic deformation parameter. At $\kappa=0$, the field ${\cal B}$ becomes sterile, and the $(2, 2)$ supersymmetry
is restored for other fields ($Z_B$ is not running at $\kappa=0$, and can be taken to be $1$).

In the $\kappa^2/g^2\to \infty$ limit, the theory reduces to the linearized model~\eqref{Llin}. The constants $\kappa$ and $\gamma$ are related as
\begin{equation}
    \gamma^2 = \dfrac{\kappa^2}{Z_B Z_\BB} \quad.
\end{equation}

Note that the Lagrangian \eqref{Lfull} does not contain $Z_A$ which can be absorbed in $g$. Thus, the bare parameters of the model are $g$, $\gamma$, $Z_B$ and $Z_\BB$. Their renormalization can be calculated either using the superfield formalism, as was done in~\cite{Cui:2011rz}, or by means of the background field method, which we briefly review in Appendix~\ref{appendix:back}.

Extending the analysis to the two-loop level gives:

\begin{alignat}{9}\label{betabtwo}
    \beta(g^2)_{\text{two-loop}} &= - \dfrac{g^2}{4\pi}
    \left[
        g^2(M+N) \left( 1+ \dfrac{\gamma^2}{2\pi} \right) -(MN+1) \dfrac{\gamma^4}{2\pi}
    \right]
    \quad,\\
    \label{betagtwo}
    \begin{split}
    \beta(\gamma^2)_{\text{two-loop}} &=
    -\dfrac{\gamma^2/(2\pi)}{1-(\gamma^2/4\pi)}
    \Bigl[
    (M+N)g^2 - (MN+1) \gamma^2
    \\&\hspace{1.97cm}+ \dfrac{\gamma^2}{8\pi} \left(
    (M+N)g^2-2(MN+1)\gamma^2
    \right)
    \Bigr]
    \quad\hspace{0.37cm}.
    \end{split}
\end{alignat}

As expected, equations \eqref{betabtwo}-\eqref{betagtwo} reproduce the results for the $\CP(N-1)$ model as one substitutes $\{M,N\}\to\{1,N-1\}$~\cite{Chen:2014efa}.

\section{Large \texorpdfstring{$N$}{N} limit \label{section:largeN}}

We now briefly discuss what we can learn from the results above regarding the large-$N$ expansion of this theory.

From \eqref{betabtwo} one can immediately read off the 't Hooft large-$N$ expansion parameter
\begin{equation}\label{tge}
    t \equiv g^2(M+N) \quad.
\end{equation}

There are a number of limits one can study:
\begin{enumerate}[label={\alph*})]
    \item $M$ fixed; $N\to\infty\s$.
    \item $\nu=M/N$ fixed, $\nu \ll 1$; $N\to\infty\s$.
    \item $\nu=M/N$ fixed, $\nu \sim 1$; $N\to\infty\s$.
\end{enumerate}

The second case 
is somewhat analogous to the Veneziano limit it QCD~\cite{Veneziano:1976wm}. In such a setting, a larger number of planar diagrams, as compared to the case of ordinary large $N$, survives. By examining~\eqref{betabtwo} and~\eqref{betagtwo}, we deduce that, in order to have a sensible limit, the second expansion parameter has to be defined as
\begin{equation}\label{tga}
    \tilde{t} \equiv \gamma^2 M N \quad,
\end{equation}
and be finite. Qualitatively, we can understand from the ${\cal B}$ one-loop propagator which has no external loops and the $MN$ degrees of freedom (indices) running in the loop and $\gamma$ at each of the vertices. The limit above makes such a diagram finite.

In this case, the beta-functions can be defined purely in terms of the t' Hooft parameters:
\begin{alignat}{9}
\label{surprisemotherfuckersONE}
\beta(t)_{\text{two-loop}}&=-\frac{t^2}{4\pi}\quad&&,\\    
\label{surprisemotherfuckersTWO}
\beta(\tilde{t})_{\text{two-loop}}&=-\frac{\tilde{t}}{2\pi}
\left(t-\tilde{t}\right)
\quad&&.
\end{alignat}
The higher-order corrections vanish in this approximation, as it can be seen from \eqref{betabtwo} and \eqref{betagtwo}, and the expressions above reduce to one loop. It should be noted that the 't Hooft limit here is defined at UV. In IR the coupling constants grow while $M$ and $N$ remain the same, so the limit is not valid anymore. 

\begin{figure}\captionsetup{font=footnotesize}
\centering
    \includegraphics[width=0.4\textwidth]{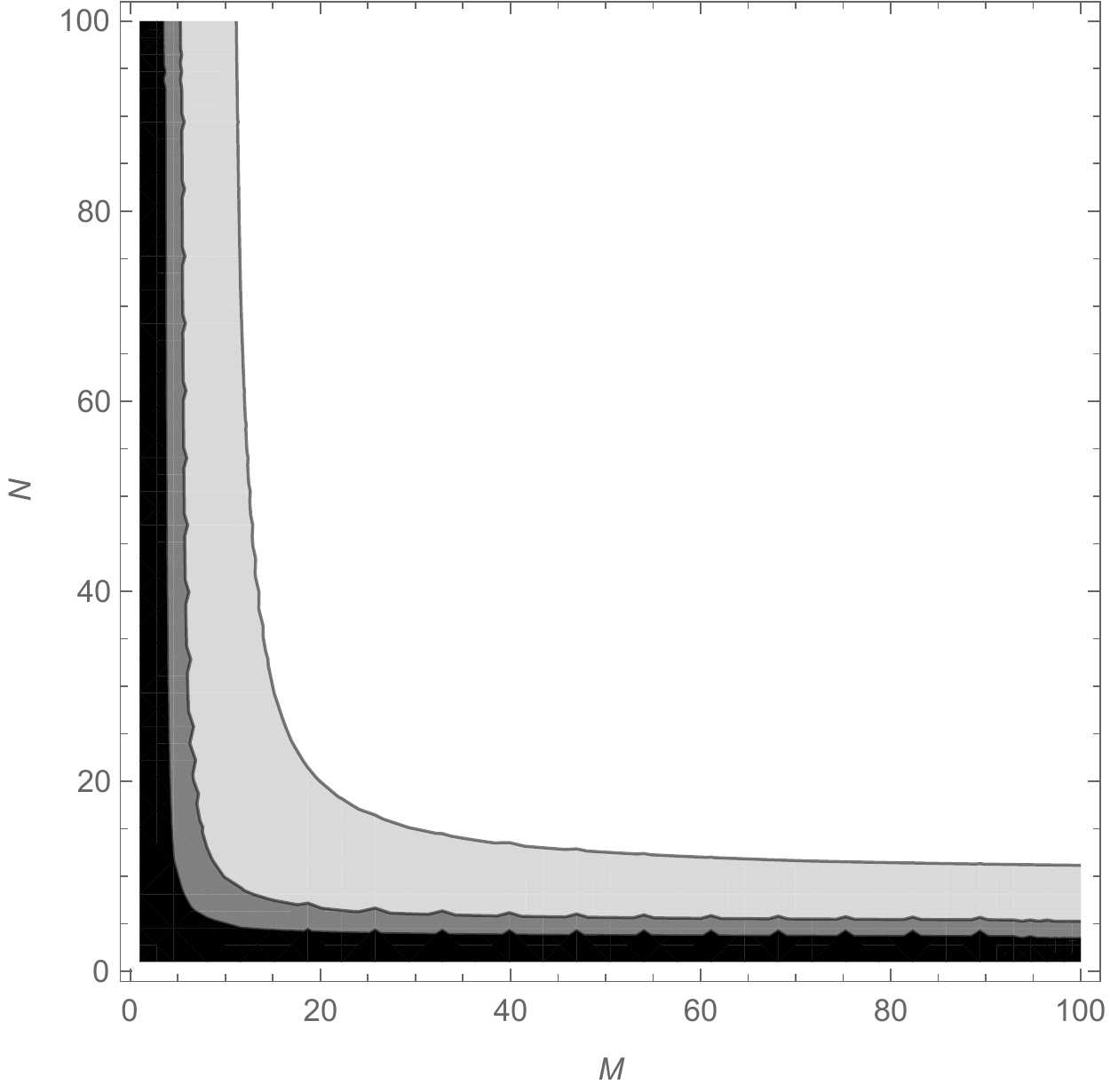}
    \includegraphics[width=0.09\textwidth]{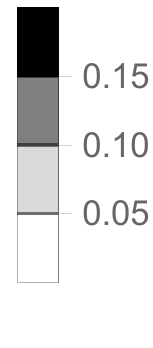} \\
  \caption{Fixed point $\rho_*$ for the ratio $\rho\equiv\gamma^2/g^2$ as a function of $M$ and $N$. The value of $\rho_*$ is denoted by brightness (the scale is on the right) while $M$ and $N$ run along both axes. We see that if one of the parameters scales with the other one, $\rho_*$ approaches 0 no matter what that ratio is.}
  \label{HeteroBetaPicture}
\end{figure}

\medskip

\begin{figure}\captionsetup{font=footnotesize}
\centering
    \includegraphics[width=0.78\textwidth]{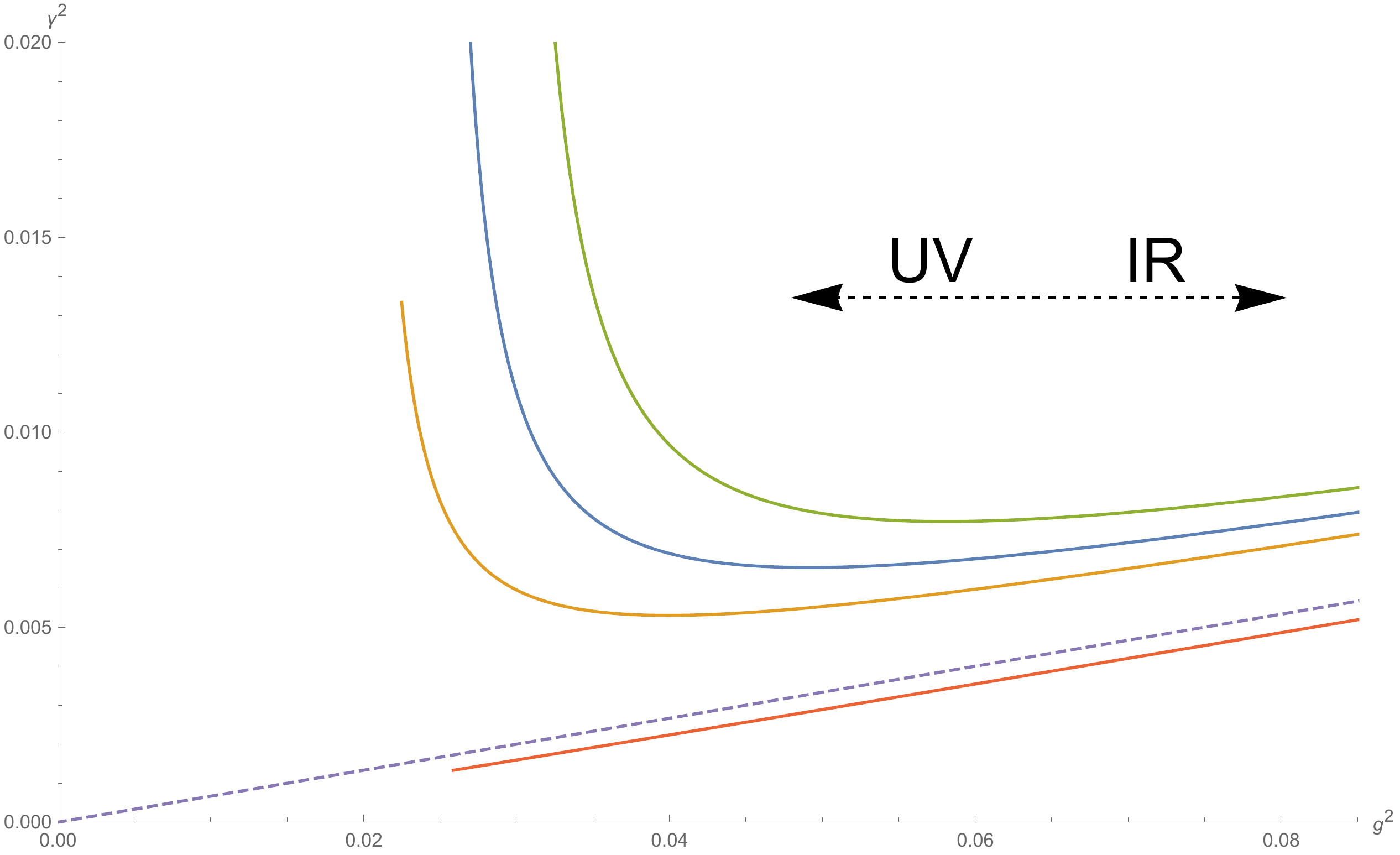}
    \caption{The RG flow of the coupling constants $g^2$ and $\gamma^2$, based on the two-loop expressions \eqref{betabtwo} and \eqref{betagtwo}. Here $M=10$, $N=30$. The dashed line has the slope $\rho_*=\dfrac{M+N}{2(MN+1)}$, so the curves above and below correspond to $\rho$ being greater and smaller than $\rho_*$ respectively. We see that the coupling constants flow to that ratio at the IR.}
  \label{RGFlowPic}
\end{figure}

What are the other consequences of considering such a limit? Let us take a look at the fixed-point behaviour of the theory. The ratio $\dfrac{\gamma}{g^2}$, being a constant in the superpotential, is not getting renormalized. However, one may be interested in the ratio
\begin{equation}
    \rho \equiv \dfrac{\gamma^2}{g^2} \quad,
\end{equation}
appearing in front of the four-fermion interaction. The corresponding one-loop beta function acquires the form of
\beq\label{betarhooneloop}
\beta(\rho)=\frac{\rho g^2}{4\pi}[2(MN+1)\rho-(M+N)]\quad.
\eeq
It has a fixed point
\begin{equation}
    \rho_* = \dfrac{1}{2} \dfrac{M+N}{MN+1} \quad .
\label{mfixed}
\end{equation}
Its behaviour as a function of $M$ and $N$ is illustrated at the Figure \ref{HeteroBetaPicture}. The fixed point \eqref{mfixed} remains at two loops, the appropriate numerical result is shown in Figure \eqref{RGFlowPic}. For $M$ being constant, this fixed point approaches the asymptotic value
\begin{alignat}{99}
    &\rho_* \longrightarrow \dfrac{1}{2M}
    \qquad \text{($N\rightarrow\infty$, $M$ fixed)} \quad.
\intertext{For $M=1$ one recovers the result $\rho_*=1/2$ for the $\CP(N)$ model ~\cite{cuiphd}.
By looking at~\eqref{mfixed} we conclude that, if $M$ scales with $N$, the fixed point value becomes}
    &\rho_* \longrightarrow 0
    \quad (N\rightarrow\infty, \varkappa \equiv\frac{M}{N} =\const\neq 0 ) \quad,
    \label{venrho}
\end{alignat}
which is the case for the Veneziano limit. Interestingly, we obtain that the value for the fixed point does not depend on $\varkappa$ once $M$ scales with $N$. There is no difference between the Veneziano limit of a small but finite $\varkappa$ and the case when $M \sim N$ and $\varkappa \sim 1$.

From \eqref{betarhooneloop} we see that for consistency we have to choose
\begin{equation}
    \rho \leq \rho_*
\end{equation}
in order to avoid the Landau pole. In the opposite case, one faces the situation similar to that in the linearized model \cite{Cui:2011rz}.

\section*{Conclusions}

In this paper, we have discussed the perturbative aspects of the heterotically deformed $(0,2)$ Grassmannian $\mathcal{G}_{M,N}$ model, as well as of its linearized version. In the latter case, the only coupling constant is the deformation parameter, whose running is determined by the renormalization of the fields solely~--- which can be proved non-perturbatively. The beta function stays positive, and the theory possesses a Landau pole in UV.

A similar kind of behaviour takes place in the full model~--- however, only for a certain range of parameters. There, in contradistinction with the linearized model, for a different set of parameters one may also end up with a well-defined asymptotically free theory, reaching its conformal point at the UV.

Lastly, we have discussed the large-$N$ and Veneziano limits of the model, which will be subject to a more detailed research in the upcoming publication~\cite{GaugedGrassmannian2}. We found that the large-$N$ behaviour of the coupling's fixed point value depends crucially on the type of limit used.

\section*{Acknowledgements}

The authors are grateful to J.~Chen both for the stimulating exchange which gave a start to this work and for the advise provided at its later stages. The authors also wish to thank X.~Cui for useful discussions on the topics raised in this paper. The work of MS was supported in part by the DOE grant \mbox{DE-SC0011842}. EK was supported in part by the Robert E Greiling Jr. Scholarship for Particle Physics and Cosmic Rays. MK was supported by DOE HEP award 0000240346.

\begin{appendices}

\section{Conventions \label{upanddicks}}
Gamma matrices, metric, $\varepsilon$-symbols:
\begin{equation}\begin{gathered}
\gamma^0 = \sigma_2 = \bpma 0 & -\iu \\ \iu & 0  \epma \quad,\qquad
\gamma^1 = \iu \sigma_1 = \bpma 0 & \iu \\ \iu & 0  \epma \quad,\\
\gamma^5 = \gamma^0 \gamma^1 = \sigma^3 = \bpma 1 & 0 \\ 0 & -1  \epma \quad.
\end{gathered}\end{equation}
\begin{equation}
\epsilon_{\alpha \beta} = \iu \sigma_2 = \pmat{0}{1}{-1}{0} \quad,\qquad
\epsilon^{\alpha \beta} = \iu \sigma_2 = \pmat{0}{-1}{1}{0} \quad.
\end{equation}
\begin{equation}
g^{\mu \nu} = \pmat{1}{0}{0}{-1} \quad,\qquad
g_{\mu \nu} = \pmat{1}{0}{0}{-1} \quad.
\end{equation}
\begin{equation}
\gamma^\mu \gamma^\nu + \gamma^\nu \gamma^\mu =2 g^{\mu \nu} \quad.
\end{equation}

\begin{equation}
\alpha,\beta = 1,2 \equiv R,L \quad,\qquad
\mu,\nu = 0,1 = t,z \quad.
\end{equation}

Spinors:
\begin{equation}
\psi = \pcolt{\psi_R}{\psi_L} \quad,\qquad
\psi^\dagger = \bpma \psi_R^\dagger & \psi_L^\dagger \epma \quad,\qquad
\overline{\psi} = \psi^\dagger \gamma^0
\end{equation}

Lightcone coordinates:
\begin{subequations}
\begin{alignat}{9}\label{defx}
x^L &= \dfrac{1}{2}(x^0 + x^1) \quad,\qquad \partial_R = \partial_t - \partial_z \quad,\\
x^R &=\dfrac{1}{2}(x^0 - x^1) \quad,\qquad \partial_L =\partial_t + \partial_z \quad.
\end{alignat}
\end{subequations}

Supercharges:
\begin{equation}
 \{ \sQ_R, \, \overline{\sQ}_R \} = - 2 \iu \partial_R  = - 2(\EuScript{H + \EuScript{P}})\quad,\qquad
 \{ \sQ_L, \, \overline{\sQ}_L \} = - 2 \iu \partial_L  = - 2(\EuScript{H - \EuScript{P}})\quad.
\end{equation}

Spinor contraction:
\begin{subequations}
\begin{alignat}{9}
&
\label{contra}
\psi \theta = \psi^\alpha \theta_\alpha = \psi^\alpha \epsilon_{\alpha\beta} \theta^\beta =  \psi_R \theta_L - \psi_L \theta_R \quad&&,
\\
&\overline{\psi}\theta = \overline{\psi} \unit \theta = \iu (\psi^\dagger_L \theta_R - \psi^\dagger_R \theta_L) \quad,\qquad
&&\overline{\psi}\gamma^5\theta =\iu (\psi^\dagger_L \theta_R + \psi^\dagger_R \theta_L) \quad&&,\\
&\overline{\psi}\gamma^0 \theta = \iu (\psi^\dagger_R \theta_R + \psi^\dagger_L \theta_L) \quad,\qquad
&&\overline{\psi}\gamma^1 \theta = \iu (\psi^\dagger_R \theta_R -\psi^\dagger_L \theta_L) \quad&&.
\end{alignat}
\end{subequations}

Integration:
\begin{equation}
(\overline{\theta}\theta) = \iu (\theta^\dagger_L \theta_R - \theta^\dagger_R \theta_L) \quad,\qquad
(\overline{\theta}\theta) (\overline{\theta}\theta) =- 2 \theta_R \theta_L \theta_R^\dagger \theta_L^\dagger \quad.
\end{equation}
\begin{subequations}
\begin{gather}
 \int \d \theta_R \, \theta_R = 1 \quad,\qquad
 \int \d \theta_L \, \theta_L = 1 \quad,\qquad
 \int \d \theta_R^\dagger \, \theta_R^\dagger = 1 \quad,\qquad
 \int \d \theta_R^\dagger \, \theta_R^\dagger = 1 \quad,\\
 \int \d \theta_R  \d \theta^\dagger_L \,(\overline{\theta}\theta) = \iu \quad,\qquad
 \int \d \theta_L \d \theta^\dagger_R \,(\overline{\theta}\theta) = \iu \quad,\\
 \int \d \theta_R \d \theta_L \d \theta_R^\dagger \d \theta_L^\dagger \,(\overline{\theta}\theta) (\overline{\theta}\theta) = 2 \quad.
\end{gather}
\end{subequations}

Chiral coordinates:
\begin{equation}
    y^\mu = x^\mu + \overline{\theta} \gamma^\mu \theta \quad.
\end{equation}

\section{Background field method\label{appendix:back}}

Following the lines for $\mathbb{CP}(N-1)$~\cite{shibook,Cui:2010si},  we start with calculating the beta-function for the Grasmannian model. The beta-function can be read off from the renormalization of the coupling constant which we will calculate using the background field method.

We begin with splitting the quantum field $\phi(x)$ in two parts:\s\footnote{~In equations~\mbox{(\ref{splitme}-\ref{justanotheruselesslabel})} it is implied that all possible indices are suppressed/contracted.
}
\begin{equation}\label{splitme}
\phi(x) = \phi_0(x) + q(x)\quad,
\end{equation}

Here $\phi_0(x)$ denotes the background field, which can be chosen arbitrarily, while $q(x)$ is the quantum correction to it.\s\footnote{~Strictly speaking, the presented approach is non-covariant from the target space point of view. Under the assumption that both $\phi(x)$ and $\phi_0(x)$ in~\eqref{splitme} belong to the target space, their difference $q(x)$ is not a well-defined geometric structure. A more careful treatment can be found in~\cite{Ketov}, where $q(x)$ is replaced by $\xi(x)$, a unit tangent vector along the target space geodesic connecting $\phi(x)$ and $\phi_0(x)$.}

We then can calculate the renormalised coupling constant by integrating out the quantum corrections to the field configuration $\phi_0(x)$:
\begin{equation}\label{effe}\begin{alignedat}{9}
    \exp\left\{-\int \d{}^2 x \, \L[g_r,\phi_r(x)]\right\}
    &\equiv \int \D\phi \, &&\exp\left\{-\int\d{}^2 x\L[g_b,\phi(x)]\right\}
    \\ &= \int \Dq \, &&\exp\left\{-\int\d{}^2 x\L[g_b,\phi_0(x) + q(x)]\right\}
    \quad,
\end{alignedat}\end{equation}
where $g_b$ and $g_r$ denote the bare and renormalised couplings, correspondingly. The RHS of~\eqref{effe} is calculated by expanding the exponent into the series in $q(x)$. As we shall see shortly, once the background field $\phi_0(x)$ has been chosen in a convenient way, the expansion of the Lagrangian acquires the form of:
\begin{equation}\label{justanotheruselesslabel}\begin{alignedat}{9}
    \L[g_b,\phi_0+q]  =  \L[g_b,\phi_0]&+
    q\times(...)
    \\&+
    C[\phi_0] \s (\partial q)^2 +V(q,\partial q) \quad.
\end{alignedat}\end{equation}
The first term is the background Lagrangian to which the quantum corrections are to be calculated. The term linear in $q$ \textit{has to vanish} within the background field method. This either happens automatically when $\phi_0$ obeys the classical equations of motion, or otherwise achieved by adding  the appropriate source terms. The third term defines the free propagator of the field $q$ ($C[\phi_0]$ is the quadratic coefficient in the Taylor expansion of $\L[g_b,\phi_0 + q]$). By calculating the loop corrections to it, we shall obtain the wave function renormalization. Lastly, by $V(q,\partial q)$ we have denoted all the remaining terms in the expansion of $\L[g_b,\phi_0+q]$. Those contain an infinite number of terms which can be represented by diagrams with $\phi_0$-dependent vertices.

For definiteness, we present the calculation of the $1$-loop beta-function of the bosonic Grassmannian model. Following the steps from Section 28 of~\cite{shibook}, we obtain, in analogy with (28.29) in \textit{Ibid.}:

\begin{equation}
    \mathcal{L}^{[2]} = 2 \partial_\mu \qb^{n\alpha} \partial^\mu q^{n\alpha} 
    - 2 k^2 |f|^2 \left(
    \sum \limits_{n=1}^{N} \qb^{n1} q^{n1} +\sum \limits_{\alpha=1}^{M} \qb^{1\alpha} q^{1\alpha}
    \right)\quad.
\end{equation}

Comparing the equation above with the $\mathbb{CP}(N-1)$ case, we deduce that:~\footnote{~$\mathbb{CP}(N-1) = \mathcal{G}_{N-1,1}$.}
\begin{equation}\label{betab}
    \beta(g^2)_{\text{one-loop}} = - \dfrac{g^4 (M+N)}{4\pi} \quad,
\end{equation}
\begin{sloppypar}
which matches the result of~\cite{Morozov:1984ad}. We recognize the dual Coxeter number ${T_{SU(M+N)}=M+N}$, which also appears in the metric, and in the Ricci tensor.
\end{sloppypar}

Next, we follow the steps of~\cite{Cui:2010si} and calculate the $Z$-factors for the heterotically deformed $(0,2)$ model. We keep using the background field method and perform the calculations in components. The only vertex relevant for the one-loop calculation is $\left(\gamma{\zeta}_RG_{i\jb\alpha\betab}(\iu\partial_L\phib^{\betab\jb}) \psi_R^{i\alpha}+\text{H.c.}\right)$.

To the lowest order, the diagram for the wave function renormalization takes the form:
\begin{equation}
\gamma^2{\zeta}_R(x)
    \delta^{i\jb}\delta^{\alpha\betab}
    \delta^{k\lb}\delta^{\gamma\deltab}
    \psi_R^{i\alpha}(x)
    (\iu\partial_L\qb^{\betab\jb}(x))
    (-\iu\partial_L q^{k\gamma}(y))
    \psib^{\deltab\lb}_{R}(y)
    \zeta_R^\dagger(y)
    \quad.
\end{equation}
For each of $\psi_R^{i\alpha}(x)$, the diagram is identical to the $\mathbb{CP}(1)$ case, while for $\zeta_R$ we are getting an additional $M\s N$ factor corresponding to the number of fields in the loop.

This way, for the $\psi_R^{i\alpha}(x)$ field one gets
\begin{alignat}{3}
    Z_\psi &= 1 + \iu \gamma^2 I \quad&&,
\intertext{while for $\zeta_R$}
    Z_\zeta &= 1 + M\s N \gamma^2 I \quad&&.
\end{alignat}
Here
\begin{equation}
    I = \int \dfrac{\d{}^dp}{(2\pi)^d} \dfrac{1}{p^2-\mu^2} = \dfrac{1}{2\pi} \log \left( \dfrac{M_{\text{uv}}}{\mu}  \right) \quad.
\end{equation}

To the first order, the bosonic beta-function~\eqref{betab} remains intact, while the beta-function for the deformation parameter is:
\begin{equation}\label{bgamma}
    \beta(\gamma)_{\text{one-loop}} = 
    \dfrac{\gamma}{4\pi} \left[
        (M+N)g^2 - (MN+1) \gamma^2
    \right]
    \quad.
\end{equation}

To proceed further, we take into account the diagrams contributing to the renormalization of $B$ and $\BB$ at two loops. %

\end{appendices}

\printbibliography[heading=bibintoc] 

\end{document}